\pageno=01
\magnification=1200 \baselineskip 18pt

\rightline{FSU-HEP-92-0818}
\hfil \break

\centerline{{\bf Asymptotic Freedom and Euclidean Quantum Gravity}
\footnote*{This research project was partially funded by the Department
of Energy under Contracts DE-FG05-87ER40319 and DE-FC05-85ER2500
and by the National Science Foundation under INT-8922411.} }
\hfill\break

\centerline{Bernd A. Berg$^{1,2}$, Balasubramanian Krishnan$^{1,2}$
            and Mohammad Katoot$^3$}
\hfill\break

\centerline{$^1$ Department of Physics,
            The Florida State University, Tallahassee, Florida 32306}
\centerline{$^2$ Supercomputer Computations Research Institute}
\centerline{The Florida State University, Tallahassee, Florida 32306}
\centerline{$^3$ Department of Mathematics and Physical Science}
\centerline{Embry-Riddle Aeronautical University, Daytona Beach, Florida 32114}
\hfill\break

\centerline{\bf Abstract}

Pure SU(2) gauge theory is the simplest asymptotically free theory in
four dimensions. To investigate Euclidean quantum gravity effects in a
fundamental length scenario, we simulate 4$d$ SU(2) lattice gauge theory
on a dynamically coupled Regge skeleton. The fluctuations of the skeleton
are governed by the standard Regge-Einstein action. From a small $2\cdot 4^3$
lattice we report exploratory numerical results, limited to a region of
strong gravity where the Planck mass and hadronic masses take similar
orders of magnitude. We find a range of the Planck mass where stable
bulk expectation values are obtained which vary smoothly with the gauge
coupling, and a remnant of the QCD deconfining phase transition is located.
\hfil \break \vfil \eject

\centerline{\bf 1. Introduction} \smallskip

Quantization of gravity is the most fundamental unsolved problem in
modern physics. Obviously, nature manages to combine gravity and the
quantum field theories of strong, weak and electromagnetic interactions
consistently, whereas all our theoretical approaches reveal serious
inconsistencies at one or the other level. Despite the fundamental
character of this problem we are in the unfortunate situation that
there are no clear-cut quantum gravity related experimental facts.
Theorists can do little more than exploring all promising branches
from a tree of alternatives
in the hope that arguments of consistency and beauty may lead to the
correct final answer. At the first look, a fairly hopeless attempt in view
of the many logical alternatives which all seem to be worthwhile to
explore. On the other hand, by now the number of theorists all around
the world is also large and, maybe, if things are done in some kind
of systematic manner one finally stumbles into the correct branch
and, hopefully, realizes it. Once a consistent, acceptable theory exists
some previously unrealized experimental evidence may also pop out.
{}From this point of view the various approaches like Supersymmetry,
String Theory, etc. all all have their own right to be pursued. Less
ambitiously, instead of constructing immediately a complete theory, one
may first study problems which arise when one tries to quantize the
classical Einstein action
$$ {\hat S}_E\ =\ m_P^2 \int \sqrt{-g}\ d^4x\ R . $$
Here $m_P = 0.17\cdot 10^{19}\ GeV$ is the Planck mass and $R$ is the
scalar Riemann curvature. Conveniently the Euclidean action,
after rotation in the complex plane [1]
$$ S_E\ =\ m_P^2 \int \sqrt{g}\ d^4x\ R , \eqno(1) $$
is used to study a variety of
problems\footnote*{In contrast to quantum field theory the usefulness of
the Euclidean rotation for quantum gravity remains hypothetical at the
present state of affairs.}. With our sign convention the Boltzmann
factor reads $\exp (+S_E)$. The conventional wisdom is that
the action~(1) describes self-interacting
spin two massless particles, but its perturbative quantization runs into
the well-known trouble that it is non-renormalizable [2] and unbounded [1].
Therefore, terms quadratic in the Riemann tensor have been introduced
which allow reformulation of gravity as an asymptotically free field
theory. However, this theory has problems with unitarity, for
references see [3].
\hfill\break

On the other hand, the possibility exists that we are confusing shortcomings
of the perturbative technique with shortcomings of the theory. A
non-perturbative approach would clearly be favourable. To pursue this
simulations are presently the only promising technique at hand.
Unfortunately, realistic quantum gravity related problems are
computationally very intensive. Even the rapid advances of modern
computer technology have just barely brought us to the point where
some exploratory studies become feasible. Quantum gravity simulation
are clearly still in their infancy. Nevertheless various directions have
already been pursued, see Menotti [3] for a review. Here we follow a line
of thought which is based on the Regge calculus [4]. The Regge calculus
replaces the smooth space-time manifold by a piecewise flat simplicial
manifold, the Regge skeleton. A $d$-dimensional simplicial lattice is
constructed by $d$-simplices
which are glued together to form a piecewise flat geometry. Every $d$-simplex
consists of ($d$+1) sites (0-simplices), $d$~($d$+1)/2 $= {d+1 \choose 2}$
links (1-simplices), ${d+1 \choose 3}$ triangles (2-simplices),
${d+1 \choose 4}$ tetrahedras (3-simplices), etc. until the dimension
$d$ is reached. On a 4$d$ simplicial manifold the action (1) becomes [4,5]
the Regge-Einstein action
$$ S_{RE}\ =\ 2 m_P^2 R , ~~~{\rm where}~~~
R\ =\ \sum_t \alpha_t A_t  \eqno(2)$$
is the scalar curvature. The sum is over all triangles (in $d$
dimensions $d-2$ simplices which are conventionally called hinges) of
the 4$d$ simplicial manifold, $A_t$ is the area of triangle $t$ and
$\alpha_t$ the associated deficit angle. The 4-simplices are pentahedra
which we label by $p$ and we denote their associated volumes by $v_p$.
The Regge skeleton is piecewise flat with the curvature concentrated on
the hinges. This enables to calculate $\alpha_t$, $A_t$, $v_p$ etc. by
the rules of elementary Euclidean geometry.
With respect to this action the Euclidean partition function is
$$ Z\ =\ \int D[\{ l\} ]\ e^{S_{RE}} . \eqno(3a)$$
Here we use $l$ in a double meaning, as a label as well as to denote the
actual link length.
The choice of the appropriate measure has been subject of much debate,
see [3] for references. Nevertheless, it is fairly unambiguously determined
for our present purposes: The dimensionful action $S_{RE}$ requires a scale
invariant measure (otherwise the dilatation $l\to l' = \lambda l$ will
already blow up the functional integral)
and the technical limitation of finite computer
speed requires a local measure. Then the ansatz
$D[\{ l\} ] = \prod_l f(l)$ implies [7]
$$ D[\{ l\} ]\ =\ \prod_l {d l \over l} . \eqno(3b) $$
Of course, this measure is not immune of critics [3] and to involve a technical
assumption cannot be regarded as satisfactory. On the other hand, the
universality assumption discussed in the next section suggests
that we may well expect physically reasonable results.
Still, to give a meaning to the dimensionful action $S_{RE}$ one has first
of all to define a scale, henceforth called fundamental length. Without
such a fundamental scale the value of $S_{RE}$ is just undefined.
By keeping one dimensional expectation value constant, most naturally
$<l>$, $<A_t>$ or $<v_p>$, the Regge calculus allows
to introduce a fundamental length~[6,7]. This matter of setting
the scale should not be confused with introducing a cosmological
constant. Once the fundamental scale is identified, all other physical
quantities may be measured in units of this scale. In particular
renormalization is no longer a problem of infinities, but just a
problem of normalization and possibly large numbers due to the fact that
the fundamental length (of the order of the Planck length) is very small
as compared to hadronic scales (of the order of a Fermi). From this
viewpoint of the Planck scale, once the natural setting has been identified,
it may well happen that normalization does not require any counterterms.
We adopt this point of view in the folowing and the cosmological constant
is then zero through the
simple fact that it is not present in the action (2). Under the
assumption that the theory exists, in the sense made precise below, every
physical quantity will be directly calculable in terms of the fundamental
units. Let us denote the fundamental length by $l_o$ and the corresponding
fundamental
mass unit by $m_0 = l_0^{-1}$. A simulation is then carried out for
$m_P = {\rm const}\ \cdot m_0$, i.e. for a Planck mass which is a number
in units
of $m_0$. It is a remarkable numerical observation [7,8] that the pure
demand of the existence of such a scale $m_0$ seems to ensure a finite range
$$ 0\ \le\ m^2_P\ <\ m^2_{MAX} \eqno(4)$$
for which the partition function (3) is well
defined. In [7] this range was called ``entropy dominated''. For a Planck mass
outside this range the partition function simply does not
exist and, therefore, it may be somewhat misleading to talk about a phase
transition (although in a similar scenario for random walks the endpoint
of the existing range exhibits indeed critical behaviour).
\hfill\break

In lack of any experimental guidance it
is a question of aesthetical simplicity to insist on the action (2).
Of course one may add terms quadratic in the Riemann tensor and imagine
that those term do not contribute in the classical limit. But new
parameters are then implied and it is difficult to imagine that these
parameters could be determined from theoretical principles alone,
although interesting suggestions, like
Weinberg's [9] proposal of gravity as an asymptotically safe theory,
exist in the literature. The reader may
consult the review [3] for work along such lines.
In this paper we embark on a suggestion of one of the authors~[10] to
explore the possible physical relevance of the entropy dominated region.
Empirically [7,8] small values for $m^2_{MAX}$ of equation~(4)
are obtained. However,
this statement is with reliance to the respective fundamental units.
The question of physical relevance is whether, within the entropy dominated
region, elementary particle masses can be chosen arbitrarily small
as compared to the Planck mass or not. In the former case this would
hint towards the possibility of a self-consistent theory of quantum
gravity, whereas in the latter case the entropy dominated region would
just be a mathematical curiousity. To investigate this problem demands
to couple gravity to matter field and to look for a point in parameter
space where the mass gap can be send to zero as compared to the Planck
mass. The issue is still subtle, as the answer may well depend on the
details of the interaction and the matter fields chosen. Here we work
under two hypothetical assumptions:

\item{i)} The world without gravity is described by a (grand unified)
          asymptotically free quantum field theory.

\item{ii)} The only properties of this quantum field theory,
           which matter for our
           questions of relative scales, are asymptotic freedom and
           the dimensionality of four space-time dimensions.

\noindent
Therefore, we decided to couple quantum gravity to the computationally
simplest 4$d$ asymptotically free field theory and this is pure SU(2)
lattice gauge theory. Let $\beta = 4/g^2$ be the coupling constant for
the SU(2) gauge theory. Within the entropy dominated range, the hope is
then to establish numerical evidence for the following scenario:

\item{a)} With increasing $\beta$, on sufficiently large systems, the
          hadronic masses (glueballs, string tension, deconfinement
          temperature) can be chosen arbitrarily small as compared to
          the Planck mass $m_P$ and, asymptotically, the perturbative
          two loop scaling formula [11] is approached as indicated in
          figure~1.

\item{b)} In the same limit the space becomes flat when averaged over
          a hadronic length scale (inverse glueball mass etc.) and,
          in particular, the Riemann curvature scalar approaches zero.

\noindent
If a) and b) hold, one could imagine that quantum gravity is simply
defined by the Regge-Einstein action in the entropy dominated region.
As indicated in figure~1, the concept of a fundamental length would then
allow us to fix $\beta$ at $\beta = \beta_{physical} = 4/ g^2_{physical}$
such that the correct value for $m_{hadronic} / m_P \approx 10^{-18}$
is obtained. In contrast, in ordinary lattice
gauge theory one would carry out the limit $\beta\to\infty$, implying
$m_{hadronic}\to 0$ in units of the lattice spacing $a^{-1}$ and employ for
$\beta=4/g^2$ the notion of a bare coupling constant. Assuming that a)
is correct, a rough order of magnitude estimate $g^2_{physical} \approx 0.5$
is obtained by equating $m_{hadronic}$ with the two loop $\Lambda_L$
scale, which in our context is conveniently defined by
$$ \Lambda_L\ =\ m_P\ \left(6\pi^2 \beta / 11 \right)^{51/121}\
                 \exp \left( -3\pi^2\beta/11 \right) . \eqno(5) $$
On the other hand, the requirements a) and b) are precise enough to allow
falsification by computer experiments. It may turn out that
the entropy dominated range is limited to $m^2_{MAX}\ \le\ const\
m^2_{hadronic}$ and~/ or that flat space cannot be approached. In some
sense the previously observed ``smallness'' of $m^2_{MAX}$ hints towards
this scenario. If this happens indeed, it would be difficult to imagine
that the entropy dominated range could be of physical relevance.
\hfill\break

The rest of this paper is organized as follows: In section~2 we define the
model and state additional assumptions. Numerical results are given in
section~3 and final conclusions can be found in section~4. Some special
problems are relegated to appendixes.
\hfill\break

\centerline{\bf 2. The Model} \smallskip

On a 4$d$ simplicial manifold the action for Regge-Einstein gravity (2)
coupled to pure SU(2) lattice gauge theory is given by
$$ S\ =\ S_{RE} + S_{gauge}, ~~{\rm where}~~
   S_{gauge}\ =\ -\ {\beta \over 2}
\sum_t W_t Re \left[\ Tr \left( 1 - U_t \right) \right] . \eqno(6) $$
As in (2) the sum goes over all triangles of a Regge skeleton and
$\beta=4/g^2$ is the gauge coupling. SU(2) matrices
$U_{ij} = U^{\dagger}_{ji}$ are associated with the links of the
skeleton, where $i, j$ are the sites of the link in question. $U_t$ is
the product of the SU(2) matrices around triangle $t$, with $i, j, k$
being the sites of this triangle:
$$ U_t\ =\ U_{ij} U_{jk} U_{ji} . \eqno(7) $$
The dimensionless
weight factors $W_t$ are functions of the link lengths and couple
the $U$'s to the geometrical structure of the skeleton. In contrast to
ordinary lattice gauge theory on the static lattice the $1$ in the term
$Tr (1-U_t)$ is now of importance due to the fact that the weight factors
$W_t$ are dynamical. Without the $1$ the lattice would be driven towards
spurious $W_t\to\infty$ configurations. As the skeleton
is piecewise flat and its curvature concentrated on the triangles, the
flat space random lattice considerations of [12] are still valid.
In the continuum limit the lattice gauge action has to converge to the
continuum Yang-Mills action. Phrased in a form appropriate for our
present purposes, this yields [13] that
$$ \sum_t W_t A^2_t\ =\ {\rm const}\ V,\ ~~~(V\
{\rm Volume\ of\ the\ lattice}) \eqno(8) $$
has to hold on each single Regge skeleton. Beyond this restriction,
and that they are functions of the link lengths, the weight factors are
arbitrary. Clearly (8) is satisfied by the choice
$$ W_t\ =\ {\rm const}\ {V_t \over (A_t)^2} \eqno(9) $$
if one assigns a 4-volume $V_t$ with each triangle such that
$$ \sum_t V_t\ =\ V \eqno(10) $$
on each skeleton.
The obviously most natural definition for $V_t$ is the {\it closest
distance definition}:  Each point in the Regge skeleton is attributed to
its nearest hinge, where the distance to a hinge is defined as the distance
to the closest point of this hinge. Up to an irrelevant sub-volume of
measure zero, every point in the skeleton is then uniquely assigned to
a hinge and for each hinge the associated volume is manifestly positive:
$$ V_t\ \ge\ 0\ ~~~{\rm for\ each}\ t. \eqno(11) $$
However, to implement the closest distance definition for simulations faces
a number of technical problems which are easiest illustrated in two dimensions.
In 2$d$ the hinges are sites and we use the generic notation $V_h$ for
the volume associated with hinges in arbitrary dimensions. The closest
distance definition leads now immediately to the dual lattice which is
constructed from the bisectors of the original lattice, see for instance [12].
Figure~2 depicts, in flat space, part of a 2$d$ random lattice and its dual.
 From the viewpoint of the $d$-simplex (triangle in 2$d$) we notice two
possibilities:

\item{1)} Its volume contributes only to the $V_h$ volumes of sites on its
          boundaries.

\item{2)} Its volume contributes also to the $V_h$ volumes associated with
          further away sites.

\noindent
It is the second case which causes troubles. Even in two dimensions we
are not aware of a possibility to implement this situation efficiently
with local formulas, as needed for simulations with finite CPU time. The
problem gets worse in higher dimensions and due to curvature (when the
geodesics through the boundary is no longer a straight line).
\hfill\break

For lattice gauge theory on a 4d random lattice Christ and Lee [14]
suggested the weight factors $W_t\ = \tau_t / A_t $ (i.e. $V_t = \tau_t A_t$),
where $\tau_t$ is the area of the dual of the triangle $t$. This choice has
been shown to fulfil (8), but it is only in 2$d$ identical with the closest
distance definition. The dual cell of a site contains all points closest
to it, but this is not true for other simplices.
For instance from our 2$d$ random lattice of figure~2
it is obvious that associating with links a volume\footnote*{We use the
notation $V_l$ because $l$ is not a hinge in 2$d$.} $V_l = l_d l /2$, where
$l_d$ is the length of the link dual to $l$, does not coincide with the
closest distance definition. Instead, applying the closest distance
definition to links in 2$d$ leads to a barycentric division of each triangle as
shown in figure~3. Similar considerations apply to the dual objects of
triangles in 4d. Furthermore, the proposal of [14] suffers from essentially
the same technical difficulties as the closest distance definition. Only
for the situation~(1) of figure~2 one has efficient closed formulas [15],
and it is tempting to extent their application to the general case as
equation (10) turns out to be still fulfilled. However, as already noted
in [15] one has then to cope with the problem that the equations may
lead to negative $V_h$ contributions when applied to the situation~(2) of
figure~2. In more details these problems are discussed in appendix~A and
the conclusion is that it is crucial to enforce the positivity condition (11).
\hfill\break

Notable are two construction which avoid negative $V_t$ values. As
discussed in [15] one may be tempted to use the approach of appendix~A and
to impose (11) by just rejecting updating steps which propose configurations
with some $V_t$ negative. This is in some analogy with implementing the
triangle inequalities and their higher dimensional generalizations
in the simulation of a Regge skeleton. Nevertheless it seems rather
unnatural to reject configurations with
perfectly well-defined geometries. The choice which is actually
implemented in the simulations reported here relies on a barycentric
subdivision of the 4-simplices which, in technical details, is outlined
in appendix~B. Similarly as with the closest distance definition, each
point in each 4-simplex gets uniquely assigned to one of the triangles
on its boundary. Figure 3 illustrates this choice for the two dimensional
case: Whereas for the links one would reproduce the closest distance
definition exactly, this is no longer the case for the sites. Their
associated volumes $V_h$ are obtained by connecting the barycenters of each
triangle with the barycenters (midpoints) of the boundary links.
\hfill\break

Our discussion of the different options to avoid negative $V_t$ contributions
leads to a fundamental problem: For discrete quantum gravity with a
fundamental length the microscopic details of the theory are supposed to
matter. How can we then possibly expect any results of physical relevance
from a fairly arbitrary construction? The answer is that the details
become relevant for true quantum gravity effect on the scale of the
Planck length. In the line of investigations proposed here we are not
aiming at calculating these effects. Instead we only like to establish
the possibility of consistency, in the sense of figure 1, of Euclidean
fundamental length quantum gravity with an asymptotically free gauge
theory. For this question we work under the conjecture:

\item{iii)} The microscopic details of the Euclidean theory of quantum
            gravity do not matter in the usual sense of {\it universality}
            in lattice gauge theory.

\noindent
This conjecture adds to the assumptions i) and ii) of the introduction.
In lattice gauge theory the microscopic details of the lattice regularization
(as well as of the action) are irrelevant for defining the correct quantum
field theory as long as one stays within one universality class. The relevant
universality class is specified by general symmetry principles and by
requesting the correct classical limit for the action, see for instance [16].
The universality principle has turned out to be a rather powerful tool as it
adds greatly to the flexibility of quantum field theory investigations,
although it may sometimes be a subject of dispute whether
two actions are in the same universality class or not. Our conjecture iii)
generalizes universality in the sense that we now assume that the microscopic
details of the Regge skeleton will at $\beta_{physical}$ only lead to
corrections which are suppressed by order $(m_{hadronic} / m_P)$ as
compared to the leading term. Of course $m_p / m_0$ ($m_0$ being the
fundamental mass unit) may greatly depend on these details, similarly
as in ordinary lattice gauge theory the ratio of $\Lambda$-scales depends
on the regularization [17].
\hfill\break

Exploiting the universality conjecture we assume now that we may define the
theory on a hypercubic lattice, without destroying the essential scaling
behaviour of figure~1. In this construction each hypercube gets divided
into 24 pentahedra (4-simplices), for more details see [10]. Further,
following [7,10] we define the fundamental length as $l_0 = (<v_p>)^{1/4}$,
i.e. by keeping the expectation value $<v_p>$ of the
volume of a pentahedron fixed. This implies that the total volume becomes
proportional to the total number of pentahedra:
$$ V\ =\ N_p (l_0)^4 . \eqno(12) $$
Obviously, we can only simulate a very tiny portion of the universe.
In our simulation the total volume is also kept constant, but in principle
one could allow for expansion or contraction of the universe by allowing
for creation and annihilation of additional simplices.
Clearly such updating steps would be fairly difficult to implement and the
scope of our investigation is restricted to link length fluctuations which
are constrained by a fixed set of incidence matrices which define the
geometry of a hypercubic lattice.
\hfill\break

Other ways to introduce the fundamental length would be to keep
$<l>$, $<l^2>$, $<A_t>$ or similar quantities fixed. It is a remarkable
observation [8] that the action stays finite with either of these three
choices. In particular it seems to be attractive to define
the fundamental length by setting the scale with link expectation values.
Pure gravity simulations with $<l^2> = {\rm const}$ have been carried out [8]
and, qualitatively, results were found rather similar to those obtained with
$ <v_p> = {\rm const}$.
\hfill\break

To conclude this section, it is certainly worthwhile to spent at least a
few thoughts on imagining qualitatively how the microscopic details of a
fundamental length quantum gravity scenario could possible look like. It
seems natural to think of sites as some kind of sources (or sinks) and of
links as some kind of flux strings. Obviously, one does not expect any
regular geometry, but has to involve some kind of random structure. The
random lattice investigations [12-14] in flat space are a good starting
point, but the fact that for gravity the links become dynamical variables
complicates matters considerably. One may have to think about re-linking
when sites (sources) come too close to one another and about implementing
transitions between different topologies [18]. However, due to the
universality conjecture these details of the fundamental length gravity theory
are argued not be important
for verifying the scaling behaviour~a) of the introduction.
In contrast, they may well be of relevance for the flatness property~b).
The reason is that (due to the conjectured
lack of renormalization) the observable
macroscopic curvature would just be the expectation value of the fluctuating
microscopic curvature. Therefore, the preferred flatness of empty space might
be related to entropy in the sense that the functional integral measure
sharply enhances the probability of flat space configurations as compared
to curved space configurations. In this connection we like to comment on
the observed [7,8] smallness of the scalar curvature expectation value
for pure entropy $m_P^2=0$ simulations, i.e. relying alone on the
measure (3b). For the hypercubic lattice the coordination numbers of
the lattice are close to those found in the average for a random lattice [7].
Still, the fixed, regularly repeating incidence matrices are clearly an
artificial constraint, just the effect might be expected to be small. For
the pure gravity case two of the authors [19] had carried out a more
detailed investigation of the action density distribution.
Triangles on the hypergeometric lattice fall in two classes: In the
first one each triangle is attached to 4 pentahedra and in
the second to 6 pentahedra, correspondingly there are topologically
distinct incidence matrices. Figure~4 depicts now the scalar  curvature
distributions constrained to each of these classes. The remarkable feature
is that one distribution is centered around a positive mean, whereas
the other is centered around a negative mean. We take this as an
indication that the small negative over-all expectation value may be
an artifact due to the hypercubic constraints and that entropy of the
true random space may favour the desired flat space. A random lattice
investigation of pure gravity would be of some interest.
\hfill\break

\centerline{\bf 3. Numerical Results} \smallskip

As we have discussed, our scale is set by adopting $l_0 = (<v_p>)^{1/4}$ as
fundamental length. This defines lattice units to which the numbers used by
the computer refer. For instance, $V=N_p$ (12) would be stored for the total
volume. We fix the normalization in equations (8), (9) and (B4) by the
choice
$$ {\rm const}\ =\ 1500 / 71 . \eqno(13) $$
This implies that in flat space
$$ {1\over 2} \langle W_t Re [ Tr (1-U_t) ] \rangle\
=\ 1 ~~~{\rm for}~~~ \beta =0 , \eqno(14) $$
and implies a convention for $\beta$ which ensures similar orders of
magnitudes for $\Lambda_L$ as typical for flat space SU(2) simulations.
Actual simulations are very CPU time intensive due to the complicated
action (6), and our present exploratory study has remained limited to a
$2\cdot 4^3$ lattice. The volume is $V=24\cdot 2\cdot 4^3 = 3,072$, as
each hypercube embeds 24 pentahedra. To compare with lattice
sizes of conventional hypercubic systems, one may equate their number
of plaquettes with the number of triangles in the present case. Conventional
lattices have six different plaquettes per hypercube, we have fifty
different triangles. This converts into a factor $(50/6)^{1/4} \approx 1.7$.
In this sense, our system is as big as a conventional one of size
$3.4\cdot (6.8)^3$.
\hfill\break

Our initial (starting)
configuration will be in flat space with SU(2) matrices
assigned randomly or ordered to the links of the system. Without
gravity updates this would just be another SU(2) lattice gauge theory
simulation on a somewhat unconventional lattice, similar to
calculations [20] done in the earlier days of lattice gauge theory.
Indeed, it has been checked [21] that this simulation gives the expected
results. In the present paper, we include now gravity with
$$ m^2_P\ =\ 0.005 \eqno(15) $$
in our lattice units\footnote*{Note that our present convention (2) for
$m^2_P$ differs by a factor of two from [7].}. This value of the Planck
mass lies within the range [7,8] for which the partition function of pure
quantum gravity simulations stays well-defined.
We have chosen the asymmetric $L_0\cdot L^3$, ($L_0 < L$)  lattice size,
because we are interested in locating remnants of the QCD deconfining phase
transition. We define the Polyakov loop $P$ in the usual way as
$P = Tr (U_1...U_n)$, where the path is closed by the periodic boundary
conditions. In our numerical calculation the product of SU(2) matrices
is only taken along hypercube edges corresponding to a straight line in
the starting configuration.
The Polyakov loop along the shortest directions ($L_0$) is regarded
as order parameter, {\it i.e.} in the limit $L\to\infty$, $L_0$ fixed,
the disordered phase is given by $<P>=0$ and the ordered phase by
$<P> \ne 0$. In the almost continuum limit $L_0\to large$, if such a
limit exists, this transition is supposed to become the
QCD (more precisely pure SU(2) gauge) deconfining phase transition. We
do not have to bother about the interpretation for small $L_0$,
essential is that the Polyakov loop stays to be an order parameter
even on strongly fluctuating systems. In the almost continuum limit
the usual interpretation is ensured, as according to b) the space
should become flat in the average too.
\hfill\break

We have simulated the system by performing alternating updating sweeps
of the gauge and the gravity action. A gauge sweep is defined by updating
each SU(2) matrix once, and a gravity sweep by updating each link length
once. Updating the gravity action will change the weight factor $W_t$
and hence influence the subsequent gauge updating. Similarly the
actual gauge configuration influences the gravity updating through the
weight factors. We simply use the notion sweep for performing a gauge
and a gravity sweep in succession. The CPU time intensive part of the
code is the gravity updating. For our $2\cdot 4^3$ system a gravity
sweep takes about 7s on an IBM 340 RISC workstation, whereas a gauge
sweep is done in 0.3s.   Our present computer resources
did not allow for finite size scaling study of the system. Therefore, we
decided to be content with a level of rigour which was typical
for pioneering lattice gauge theory studies [23], and to employ the method
of thermal cycles for gaining an idea of the phase structure. After each
$N_1$ sweeps we vary $\beta $ by
$$ \triangle \beta\ =\ (\beta_{max}-\beta_{min})/N_2. \eqno(16) $$
We start off with $\beta = \beta_{min}$ and perform $N=N_1\cdot N_2$ sweeps,
increasing $\beta$ by $\triangle\beta$ after each $N_1$ sweeps.
Then we reach $\beta_{max}$ and perform another $N$ sweeps,
decreasing now $\beta$ by $\triangle\beta$ after each $N_1$ sweeps.
At each $\beta$ value thermal averages over the $N_1$ sweeps for various
physical quantities are measured.
For a suitable choice of $\beta_{max}$, $\beta_{min}$ and $N_1$, $N_2$
a phase transition will then show up as a hysteresis in the time series of
one, some or all of these quantities.
\hfill\break

Here we present results for the following quantities: the Polyakov loop
$P$ (closed in the $L_0=2$ direction, the gauge action $S_{gauge}$ and
$Tr (U_t) / 2$, the link lengths $l$, the triangle areas $A_t$ and the
weight factors $W_t$, last not least for the deficit angle $\alpha$ and
the gravity action $S_{RE}$, which is of course equivalent to the
scalar curvature $R$. The normalization of the gauge and gravity
actions is per triangle, for all other quantities the normalization is
according to their definitions.
Note that in contrast to flat space there is no trivial one-to-one
correspondence between $S_{gauge}$ and $Tr (U_t)$ anymore, due to
the fact that the weight factors $W_t$ have become dynamical. Also
remember that our definition of the gauge action (6) includes the
factor $-\beta/2$.
To locate the transition in question $\beta_{min} = 1.3253$,
$\beta_{max} = 1.635$ $N_1 = 300$ and $N_2 = 36$ have turned out to be
suitable parameter values. Before going through the thermal cycle, we
performed 20,000 sweeps at $\beta_{min}$. This has turned out to be
relevant due to slow equilibration of the gravity part.
Figure 5 shows hysteresis behaviour of the Polyakov loop, indicating the
existence of a phase transition. As illustrated by figures 6 and 7, the
transition shows also up in other gauge quantities such as $Tr (U_t)$ and
the gauge action itself. Further, it is noticeable in geometrical quantities,
such as the link lengths, triangle areas and
the weight factors, see figures 8, 9  and 10.
However, in the gravity action itself it is obscured by the noise of
the deficit angle, see figure 11 and 12.
Together these graphs present convincing evidence for a phase transition
around $\beta \approx 1.5$. It would be premature to comment on the
order of this transition. It is kind of interesting that the transition
does not only show up in the gauge part, but also in geometrical quantities.
However, one should not jump into speculations. If the outlined physical
picture of an almost continuum limit is correct, one expects the effects
in the geometry to become suppressed by many orders of magnitude. It is
only due to our extreme strong coupling limit that gravity and hadronic
scale take on the same order of magnitude.
\hfill\break

After having located the transition, we performed longer runs to probe
the broken phase at selected $\beta$ values: $\beta = 1.5147$, 1.5265
and 1.5383. Each run consists
of 64,000 sweeps, where an additional 20,000 initial sweeps are omitted
for reaching thermal equilibrium. Averaged over the lattice,
Polyakov loop distributions are depicted in figures 13.
Lattice averages are taken. With increasing $\beta$ a double peak
structure develops which is typical for the ordered phase.
The existence of the phase transition is therefore clear. To determine
its location more precisely would require to employ finite size scaling
techniques ($L=8, 16, ...$), but our present computer resources do not
allow for this. Even less they allow for the most interesting increase
of $L_0$. Instead, we intend to study first the phase diagram in the
$m^2_P-\beta$ plane, a task possible within our present limitations.
To provide some reference, table~1 collects from our long runs
the expectation values of the various physical quantities which we
have considered.
\hfill\break

\centerline{\bf 4. Conclusions} \smallskip

Simulating Regge-Einstein euclidean gravity coupled to pure SU(2) lattice
gauge theory on a small $2\cdot 4^3$ lattice, we find that an ``entropy
dominated'' region still exists where stable bulk expectation values
are obtained. In addition, our data support the existence of a phase
transition between a disordered and an ordered phase, the Polyakov
loop being the order parameter.
However, we cannot yet present evidence for property a)
of the scenario outlined in the introduction. To provide modest evidence,
would require to simulate a $4\cdot 8^3$ system. Keeping the
Planck mass $m_P$ fixed (in the fundamental units), one likes to find
that one is still in the well-defined region and that (in the same units)
the critical temperature moves to a smaller value
$T_c\ (L_t=4)\ <\ T_c\ (L_t=2)$. Concerning property b), we would like to
observe a decline of $<R>$, but finally microscopic details of the gravity
action may matter for this quantity.
Presently, the major stumbling block against simulating the $4\cdot 8^3$
system is lack of computer power. Already
for the $2\cdot 4^3$ system we needed considerable computing resources
and with our present allocation we could not afford a factor $> 16$ as
required for simulating the
$4\cdot 8^3$ system. Assuming continuing rapid improvements of
computer technology these simulation should, however, be feasible within
the next few years.
\hfill\break

{\bf Acknowledgements}

The Monte Carlo data were produced on FSU's CRAY-YMP, and on the SCRI
cluster of RISC workstations.
\hfil \break

\centerline{\bf Appendix A} \smallskip

In this appendix we summarize problems encountered with implementing the
dual construction locally. We consider 2$d$ where the dual vertex is located
by the intersection of the bisectors of the links. Figure~14 shows the
the case (1) discussed in section 2. The intersection of the bisectors is
inside the triangle, and exact formulas for the
contributions to the vertices can be calculated in terms of the link
lengths. For example, the contribution to A inside triangle
$ABC$ is given by the sum of area $ADG$ and $AGF$:
$$ V_A\ =\ \left( {\hat{e}_z \over 2} \right) \cdot \left( \vec{AD}
\times \vec{AG} + \vec{AG} \times \vec{AF} \right) .   \eqno(17) $$
Here $\hat{e}_z$ is the unit vector perpendicular to the plane of the triangle.
This formula and 4d versions, as needed for the proposal of Christ and
Lee [14], can be found in [15]. However, for case (2) of section 2 there
are instances  when these formulas fail to yield the area closest to the
vertex. This is a consequence of applying the local formulas to situations
where more complicated considerations are required. In particular hinge
contributions can even become negative, as in the example [15] described now.

Consider figure~15 where the perpendicular bisectors intersect at a point
$G$ outside the triangle. The area closest to $A$ is $AHF$, whereas
equation (17) gives the area $AGF$ minus the area $AGD$. The contribution
from $AGF$ is negative, because the orientation of $\vec{AG} \times \vec{AF}$
has changed. In the case of the figure one clearly sees: area $AGF >$
area $AGF$, hence the summed up contribution is negative. Still,
equation (17) has an attractive feature: One easily verifies that
the sum of all contributions $V_A+V_B+V_C$ is identical to the area of the
triangle $ABC$. Therefore, and as it allows for a straightforward
computer implementation, we could not resist to try it out. One might
imagine constraints such that large negative contributions imply
competing large positive contributions. We performed simulations with
the $4d$ analogues of formula (17). The numerical result is that reasonable
bulk expectation values are never obtained. The system builds up
configuration with large negative contributions, obviously by having
$\tau_t$ negative and $A_t$ small in the equation $W_t = \tau_t / A_t$,
and never ever escapes from there (terms of order five and more in the
exponent are typical magnitudes).

Figure~16 depicts a test run at $m_P^2 = 0.005$ for the relatively
strong coupling $\beta = 710/1500$. One sees the gauge action rendered
unbounded from above, catalyzed by the weight factors which acquire
negative contributions. The system jumps from one plateau to the next,
each time increasing its action, and cannot get out again.
Each plateau indicates a time preriod when nearly all
new proposed link lengths were rejected. We conclude that the requirement
(11), $V_t \ge  0$, is crucial for numerical simulations of the system.
\hfill \break \vskip 15pt

\centerline{\bf Appendix B} \smallskip

Here we give in details the barycentric definition of the volumes $V_t$
as used when implementing equation (9) for our present simulations.
Figure~17 illustrates the barycentric subdivision of the volume of a
typical pentahedron $ABCDE$. We consider the construction of the
sub-volume which adds to the volume associated with the triangle $ABC$.
We proceed by first locating the barycenters of the pentahedron and
that of the two tetrahedra that share the triangle $ABC$. Let the position
vectors of the vertices A, B, C , D, and E be $\vec{0}$, $\vec{r_1}$,
$\vec{r_2}$, $\vec{r_3}$, and $\vec{r_4}$ respectively.
The barycenter of the pentahedron $ABCDE$ is then given by
$$ \vec{G}\ =\ {1\over5} \left( \vec{0}+\vec{r_{1}}+\vec{r_{2}}+
             \vec{r_{3}}+\vec{r_{4}} \right) , \eqno(18) $$
the barycenter of the tetrahedron  $ABCD$ is
$$ \vec{F}\  =\ {1\over 4} \left( \vec{0}+\vec{r_{1}}+\vec{r_{2}}+
              \vec{r_{3}} \right) , \eqno(19) $$
the barycenter of the tetrahedron $ABCE$ is
$$ \vec{F'}\ =\ {1\over 4} \left( \vec{0}+\vec{r_{1}}+\vec{r_{2}}+
              \vec{r_{4}} \right) . \eqno(20) $$
The 4-volume in the pentahedron $ABCDE$ which is now associated with the
triangle $ABC$ the combined 4-volumes of the sub-pentahedra $ABCFG$
(constructed in figure B1) and $ABCF'G$.
These contribution are manifestly positive as the barycenter lie always
inside the pentahedra. Further, the sum of the contributions to all
ten triangles of a pentahedron adds up to the 4-volume of
the pentahedron. The weight factor of triangle $t$ from the
barycentric subdivision is consequently
$$ W_t\ =\ {\rm const}\ \sum_{p \owns t} {V_t (p) \over (A_t)^2 }\
        =\ {\rm const}\ {V_t \over (A_t)^2} , \eqno(21) $$
where the sum is over all pentahedra which contain $t$ and $V_t (p)$
is the sub-volume of pentahedra $p$ which is associated with $t$.

\hfil \break
\vfill\eject

\centerline{\bf References}  \hfil\break

\item{1)} S.W. Hawking, in {\it General Relativity - An Einstein centenary
          survey}, S.W. Hawking and W. Israel (Eds.), Cambridge University
          Press 1979, p.746 ff.

\item{2)} S. Deser and P. van Nieuwenhuizen, Phys. Rev. D10 (1974) 401;
          G. 't Hooft and M.~Veltman, Ann. Inst. H. Poincar\'e 20 (1974) 69.

\item{3)} P. Menotti, Nucl. Phys. B (Proc. Suppl.) 17 (1990) 29.

\item{4)} T. Regge, Nuovo Cimento 19 (1961) 558.

\item{5)} H. Cheeger, W. M\"uller and R. Schrader, Comm. Math. Phys.
          92 (1984) 405; R.~Friedberg and T.D. Lee, Nucl. Phys. B242
          (1984) 145; M. Rocek and R.M. Williams, Z.~Phys. C21 (1984) 371.

\item{6)} G. Feinberg, R. Friedberg, T.D. Lee and H.C. Ren, Nucl. Phys.
          B245 (1984) 343; T.D. Lee, {\it Discrete Mechanics}, in Proceedings
          of the 1983 Erice Summer School, A.~Zichichi (Ed.), Plenum Press.

\item{7)} B.A. Berg, Phys. Rev. Lett. 55 (1985) 904 and
          Phys. Lett. 176B (1986) 39.

\item{8)} W. Beirl, E. Gerstenmayer and H. Markum, {\it Exploration of
          simplicial gravity in four dimension}, preprint, TU Wien 1991.

\item{9)} S. Weinberg in {\it General Relativity - An Einstein centenary
          survey}, S.W. Hawking and W. Israel (Eds.), Cambridge University
          Press 1979, p.790 ff.

\item{10)} B.A. Berg, in {\it Particle Physics and Astrophysics}, Proceedings
           of the XXVII Int. Universit\"atswochen f\"ur Kernphysik,
           H. Mitter and F. Widder (Eds.), Springer, 1989, p.223 ff.

\item{11)} W.E. Cashwell, Phys. Rev. Lett. 33 (1974) 244; D.R.T.
           Jones, Nucl. Phys. B75 (1974) 431.

\item{12)} N.H. Christ and T.D. Lee, Nucl. Phys. B202 (1982) 89

\item{13)} N.H. Christ and T.D. Lee, Nucl. Phys. B210 (1982) 310.

\item{14)} N.H. Christ and T.D. Lee, Nucl. Phys. B210 (1982) 337.

\item{15)} H.W. Hamber and R.M. Williams, Nucl. Phys. B248 (1984) 392.

\item{16)} N.S. Manton, Phys. Lett. 96B (1980) 328.

\item{17)} A. Hasenfratz and P. Hasenfratz, Nucl. Phys. B193 (1981) 210,
           and references given therein.

\item{18)} R. Geroch and J. Hartle, Foundations of Physics 16 (1986) 533.

\item{19)} B.A. Berg and B. Krishnan, unpublished (1989).

\item{20)} W. Celmaster and K.J.M. Moriarty, Phys. Rev. D33 (1986) 3718.

\item{21)} B. Krishnan, Ph. D. thesis, in preparation.

\item{22)} M. Creutz, L. Jacobs and C. Rebbi, Phys. Rev. D20 (1979) 1915.
\hfill\break \vskip 20pt

\centerline{\bf Table 1 } \hfil\break
Expectation values for $Tr (U_t) / 2$, $S_{gauge}$, $l$, $A_t$, $W_t$,
$\alpha$ and $R$. Each measurement relies on 64,000 sweeps and
measurements after spending 20,000 sweeps for reaching thermal
equilibrium. Error bars (in parenthesis) refer to the last digit.
$$\vbox{\settabs 8 \columns
\+        &         &           &        &       &        &       &     \cr
\+ $\beta$&$Tr(U_t)/2$&$S_{gauge}$& $l$  & $A_t$ & $W_t$& $\alpha$& $R$ \cr
\+        &         &           &        &       &        &       &     \cr
\+ 1.5147 &0.399(3) &$-$0.535(1)&3.341(4)&4.29(1)&0.676(2)&0.0340(6)
                                                          &$-$0.411(5)\cr
\+ 1.5265 &0.416(3) &$-$0.528(1)&3.333(4)&4.27(1)&0.683(2)&0.0346(7)
                                                          &$-$0.407(8)\cr
\+ 1.5383 &0.442(4) &$-$0.518(1)&3.310(8)&4.22(1)&0.695(2)&0.0308(9)
                                                          &$-$0.399(4)\cr
\+        &         &           &        &       &        & &         \cr}$$
\hfill\break \vskip 20pt

\hfill\break \vfill
\bye \end